\documentclass[sigconf,screen]{acmart}

\usepackage{graphicx} 
\usepackage{cite}
\usepackage{hyperref}
\usepackage{subfiles}
\usepackage{algorithm}
\usepackage{algpseudocode}
\usepackage{listings}
\usepackage{xcolor}
\usepackage{multirow}
\usepackage{multicol}
\usepackage{longtable}

\def\BibTeX{{\rm B\kern-.05em{\sc i\kern-.025em b}\kern-.08em
    T\kern-.1667em\lower.7ex\hbox{E}\kern-.125emX}}

\begin{abstract}
Large Language Models (LLMs) have achieved remarkable success in automated code translation. While prior work has focused on improving translation accuracy through advanced prompting and iterative repair, the reliability of the underlying evaluation frameworks has received less attention. In this paper, we demonstrate that a significant number of reported failures in code translation are not due to incorrect logic, but rather \textit{evaluation-induced errors} stemming from improper compilation flags, missing library links, and unconfigured runtime environments. We conduct a large-scale empirical study across five programming languages (C, C++, Java, Python, Go) and three benchmarks (Avatar, CodeNet, EvalPlus), covering 6,164 translations generated by GPT-4o, DeepSeek-Coder, and Magicoder. Our analysis identifies and categorizes common false negatives, distinguishing pipeline-induced failures that affect any model from model-dependent behaviors that vary across LLMs. Our findings highlight the necessity for transparent, configuration-aware evaluation standards to accurately assess progress in LLM-based code translation.
\end{abstract}


\title{Beyond Translation Accuracy: Addressing False Failures in LLM-Based Code Translation}

\author{Fazle Rabbi}
\orcid{0009-0007-8992-9682}
\affiliation{%
  \institution{Concordia University}
  \city{Montreal}
  \country{Canada}
}
\email{fazle.rabbi@mail.concordia.ca}

\author{Soumit Kanti Saha}
\orcid{0009-0000-8771-7829}
\affiliation{%
  \institution{Concordia University}
  \city{Montreal}
  \country{Canada}
}
\email{s_soumit@live.concordia.ca}

\author{Jinqiu Yang}
\orcid{0000-0003-4282-406X}
\affiliation{%
  \institution{Concordia University}
  \city{Montreal}
  \country{Canada}
}
\email{jinqiu.yang@concordia.ca}

\setcopyright{cc}
\setcctype{by}
\acmDOI{10.1145/3805760.3814890}
\acmYear{2026}
\copyrightyear{2026}
\acmISBN{979-8-4007-2601-9/2026/07}
\acmConference[AIware '26]{Proceedings of the 3rd ACM International Conference on AI-Powered Software}{July 6--7, 2026}{Montreal, QC, Canada}
\acmBooktitle{Proceedings of the 3rd ACM International Conference on AI-Powered Software (AIware '26), July 6--7, 2026, Montreal, QC, Canada}
\acmSubmissionID{fseaiware26main-pp054-p}
\received{2026-02-15}
\received[accepted]{2026-03-28}


\begin{document}

\maketitle

\section{Introduction}
Code translation, migrating software from one programming language to another while preserving functional correctness, is a long-standing challenge in software engineering. Such migration is critical for maintaining legacy systems~\citep{zhang2023ownership, ling2022rust, emre2021translating}, improving performance~\citep{sneed2010migrating, gandhi2024translation}, and ensuring long-term maintainability~\citep{seacord2003modernizing}, yet it often requires extensive manual effort and expert knowledge. Recent advances in large language models (LLMs) have significantly improved automated code translation~\citep{pan2024lost}, making it feasible to translate non-trivial programs across languages such as C, C++, Java, Python, and Go.

Despite these advances, fully automated code translation remains unreliable in practice. Prior work~\citep{pan2024lost, macedo2024intertrans, rabbi2025babelcoder} has largely evaluated translation quality using compilation success and test case execution as the primary indicators of correctness. While this methodology is reasonable in principle, it implicitly assumes that the evaluation pipeline, including compilation commands, flags, and runtime configurations is itself correct and faithful to the translated program's requirements.

In this paper, we show that this assumption often does not hold. Through an in-depth analysis of existing code translation artifacts and benchmarks, we find numerous cases where translated code is \emph{semantically correct} yet reported as incorrect due to evaluation mis-configurations such as missing compiler flags and incorrect linker commands. In particular, missing or incorrect compiler flags (e.g., omitted libraries, language standards, or optimization constraints) can lead to compilation or runtime failures that are unrelated to translation quality.

This issue is especially problematic in large-scale empirical studies, where evaluation pipelines are automated, and errors introduced by incorrect compilation settings propagate silently into reported results. As a consequence, translation systems may be unfairly penalized, and empirical conclusions about model performance may be misleading.

To systematically study this problem, we design an analysis pipeline that inspects translation and evaluation outcomes across three benchmarks, five languages, and 6,164 translations. By separating evaluation-induced errors from genuine translation errors, our analysis identifies a significant class of false negatives in existing benchmarks.

\noindent In summary, this paper makes the following contributions:
\begin{itemize}
    \item We conduct a systematic, large-scale analysis of failure cases in LLM-based code translation across three widely used benchmarks and five programming languages, covering 6,164 translations.
    \item We identify and categorize concrete sources of evaluation-induced false failures, including missing compilation flags, incorrect linker commands, and unrealistic timeout constraints, as well as model-dependent behaviors such as faulty test conversion and uninvoked entry-point logic.
    \item We distinguish \textit{pipeline-induced} failures, which affect any translation model under the same evaluation setup, from \textit{model-dependent} failures whose frequency varies across LLMs.
    \item We analyze results across GPT-4o, DeepSeek-Coder, and Magicoder, confirming that pipeline-induced issues are not artifacts of a specific model.
\end{itemize}

\section{Related Work}
LLM-based code translation has been widely studied at the function and file levels. Prior work improves translation quality through architectural adaptations and contextual augmentation. For example, SteloCoder~\citep{pan2023stelocoderdecoderonlyllmmultilanguage} adapts decoder-only multilingual models for code translation, while Spectra~\citep{nitin2024spectraenhancingcodetranslation} and Saha et al.~\citep{saha2024specification} leverage natural language or structural specifications as intermediate guidance. Retrieval-augmented approaches further improve few-shot translation by injecting relevant examples~\citep{bhattarai2024enhancing}, and pseudocode-based methods translate abstract representations into executable code~\citep{yu2023pseudocode}.

Several works focus on correcting translation errors after generation. Rectifier~\citep{yin2024rectifiercodetranslationcorrector} learns from faulty--corrected code pairs, while UniTrans~\citep{yang2024exploring} and ExeCoder~\citep{he2025execoder} use execution feedback and semantic information to iteratively repair translations. Lost in Translation~\citep{pan2024lost} categorizes common bugs and failure modes in LLM code translation. Multi-agent systems such as TransAgent~\citep{yuan2024transagentllmbasedmultiagentcode} and BabelCoder~\citep{rabbi2025babelcoder} distribute responsibilities across translation, testing, and repair agents. These methods primarily target solving translation-induced errors while translating code.

Repository-level translation has also been explored. K3Trans~\citep{ou2025enhancing} incorporates dependency and historical knowledge, while AlphaTrans~\citep{ibrahimzada2024alphatrans} applies static analysis and neuro-symbolic decomposition. RepoTransBench~\citep{wang2024repotransbench} provides a benchmark highlighting challenges in translating real-world repositories with automated tests.

Some recent works on code generation emphasize synthesizing executable code from natural language instructions, docstrings, or functional specifications rather than translating from existing implementations. In single-agent approaches, models such as CatCoder~\citep{pan2024enhancing} and TOOLGEN~\citep{wang2024teaching} enhance generation by incorporating structural repository context or mimicking developer-tool interactions. Multi-agent frameworks further distribute these tasks; for instance, Self-collaboration~\citep{dong2024self} and CodeAgent~\citep{zhang2024codeagent} simulate software team dynamics or integrate external tools to improve reliability. While these methods share architectural similarities with translation, particularly when intermediate specifications are involved, they rely on sparse natural language intent.

Beyond translation, broader studies on LLM-generated code have examined reliability from multiple angles, including robustness 
across programming languages~\citep{rabbi2025multi,rabbi2026hej}, security vulnerabilities in generated code~\citep{li2026exploratory,li2025secure,li2025prompt, cheng2025cfceval}, and bias in LLM outputs~\citep{ling2025bias, rabbi2026socialbias}. These findings collectively highlight that evaluation reliability and output trustworthiness remain open challenges across LLM-based code tasks.

Unlike prior work, which targets translation quality, we focus on evaluation reliability, specifically, how flawed evaluation pipelines cause semantically correct translations to be misclassified as failures.

\section{Methodology}

We analyze false failure cases in LLM-based code translation using an automated translation and evaluation pipeline. Given a source program and target language, the pipeline (1) translates the code using an LLM, (2) compiles and executes the translated program using dataset-provided test cases, and (3) records compilation errors, runtime failures, and test mismatches.

\subsection{Inspection Process}

Our study covers 6,164 translated programs across the Avatar, CodeNet, and EvalPlus benchmarks, of which approximately 600 initial failures were observed from GPT-4o. To investigate failure causes, we manually inspected 150 sampled failure cases across datasets and target languages generated by that LLM. For each case, the inspection followed a structured process: we first examined compilation logs, then runtime traces and test outputs, and finally analyzed the evaluation configuration, including compiler flags, library linking, test conversion, and runtime settings, while keeping the translated code unchanged. Three annotators each reviewed separate, non-overlapping subsets; inter-rater agreement was therefore not computed.

\subsection{Classification Criteria}

A failure is classified as \textit{evaluation-induced} if correcting the evaluation configuration alone, without modifying the translated code, allows the program to compile and pass the benchmark tests. Ambiguous cases were resolved by adjusting individual configuration components in isolation to identify the source of failure.

We distinguish two categories of non-genuine failures throughout this paper. \textit{Pipeline-induced} failures originate from the evaluation environment itself, such as missing linker flags, incorrect compilation commands, or unrealistic timeout limits, and would affect any translation model evaluated under the same pipeline. \textit{Model-dependent} failures may vary in frequency across LLMs; these include natural language contamination in generated output, excessive global memory allocation, faulty PyTest-to-JUnit test conversion, and uninvoked entry-point logic. These cases are reported separately in the Findings and discussed further in Section~\ref{sec:threats}.

This pipeline is used solely as an analysis tool to diagnose failure modes in existing benchmarks, not as a new translation or repair system.

\subsection{Applying Fixes Across Models} After inspecting the 150 failure cases produced by GPT-4o and identifying their causes, we applied the corresponding fixing strategies to the generated codes by DeepSeek-Coder and Magicoder, and report the results in Table~\ref{tab:model_dependent} for the model-dependent output issues. 

\section{Experimental Setup}

We focus on five programming languages: C, C++, Java, Python, and Go. We use three benchmarks following the same sampling as Pan et al.~\citep{pan2024lost}: Avatar~\citep{ahmad-etal-2021-avatar} (Python and Java; 250 samples per translation direction), CodeNet~\citep{puri2021codenet} (Python, Java, C, C++, and Go; 200 samples per language pair), and EvalPlus~\citep{liu2024your} (Python to Java; 164 samples). Avatar and CodeNet are translated to all five target languages, while EvalPlus targets Java only. This results in 6,164 total translations, among which approximately 600 initial failures were observed. Inspection proceeded in three stages: compilation errors were analyzed first, followed by assertion failures, and finally runtime and timeout issues.

We run all translations using three models: GPT-4o, DeepSeek-Coder~\citep{guo2024deepseek}, and Magicoder~\citep{wei2023magicoder}. Translated programs are compiled and executed using dataset-provided test cases via command-line execution. For each run, we record compilation outcomes, runtime behavior, and test results.

\section{Findings}
\label{sec:findings}

Our analysis of 150 manually inspected failure cases reveals three categories of non-genuine failures: \textit{pipeline-induced} errors that would affect any translation model under the same evaluation setup, \textit{model-dependent} errors whose frequency varies across LLMs, and \textit{genuine translation limitations} that reflect the inherent difficulty of cross-language semantics. We report these separately to avoid conflating evaluation pipeline flaws with model limitations.

\subsection{Pipeline-Induced Compilation Errors}

\noindent\textbf{Missing library linking.}
Automated evaluation scripts typically rely on generic, minimal compilation commands (e.g., \texttt{gcc file.c -o file}), which are often insufficient for complex programs. Translated C programs frequently require explicit linker flags to resolve external dependencies. Common examples include \texttt{-lm} for mathematical functions in \texttt{math.h}, \texttt{-lgmp} for arbitrary-precision arithmetic in \texttt{gmp.h}, and \texttt{-lssl} / \texttt{-lcrypto} for cryptographic operations in \texttt{openssl/ssl.h}. The omission of these flags leads to linker errors that are incorrectly classified as translation failures. We suggest that evaluation pipelines detect missing dependencies at setup time and report them separately from translation failures.

\subsection{Language-Specific Compilation Behavior}

The nature of compilation failures exhibited distinct patterns depending on the target language's strictness and paradigm. 

\begin{itemize}
    \item \textbf{Python}: We observed virtually no compilation-stage errors. As an interpreted language, Python's failures are predominantly runtime-based.
    \item \textbf{Java/C/C++}: These statically typed languages suffered mostly from syntax violations and missing import statements. In C++, template instantiation errors and header dependency chains were frequent sources of build failure.
    \item \textbf{Go}: The Go compiler enforces strict discipline regarding unused imports and variables. LLMs often translate logic directly from looser languages without cleaning up redundant variable declarations, causing frequent build failures unique to Go's strict tooling.
\end{itemize}

\subsection{Model-Dependent Output Issues}

The following issues may arise from model behavior rather than evaluation settings. Because occurrence rates are non-deterministic and sensitive to prompt phrasing, we report frequency using a qualitative scale: \textit{Low} ($<$1\%), \textit{Medium-Low} ($\sim$5\%), \textit{Medium} ($\sim$10--20\%), \textit{Medium-High} ($\sim$20--40\%), and \textit{High} ($>$40\%). Table~\ref{tab:model_dependent} summarizes these ratings across the four model-dependent categories.

\begin{table}
\centering
\footnotesize
\caption{Model-dependent failure frequency across three models. Ratings: Low ($<$1\%), Medium-Low ($\sim$5\%), Medium ($\sim$10--20\%), Medium-High ($\sim$20--40\%), High ($>$40\%). $^\dagger$Few problems in the datasets trigger this pattern; when they do, all models consistently reproduce the source-language behavior.}
\label{tab:model_dependent}
\begin{tabular}{|p{2.5cm}|p{1.5cm}|p{1.5cm}|p{1.5cm}|}
\hline
\textbf{Category} & \textbf{GPT-4o} & \textbf{DeepSeek-Coder} & \textbf{Magicoder} \\
\hline
Natural language contamination    & Low        & Low        & Medium-Low \\
\hline
Excessive global memory usage$^\dagger$ & Low   & Low        & Low        \\
\hline
Faulty PyTest-to-JUnit conversion & Medium-Low        & Medium-Low        & Medium        \\
\hline
Uninvoked entry-point logic       & Medium-Low        & Medium-Low        & Medium-Low        \\
\hline
\end{tabular}
\end{table}

\noindent\textbf{Natural language contamination.}
LLMs wrap code blocks in Markdown formatting (e.g., \texttt{```cpp}). They also frequently intersperse natural language explanations or conversational fillers within the generated source code, such as ``here is the solution'', even if they are instructed to return only the code, they return the response with natural language occasionally. Also, different LLMs use different end and start tokens that wrap the code. If these artifacts are not sanitized before compilation, standard compilers interpret them as syntax errors, causing functionally correct code to be rejected. When the prompt explicitly instructs the model to generate code without any natural language or description, GPT-4o and DeepSeek-Coder produce contaminated output in fewer than 1\% of cases (Low). Magicoder is more prone to mixing natural language with code and uses different formatting conventions, resulting in a Medium-Low rate of approximately 5\%.

\noindent\textbf{Excessive global memory usage.} 
We observed instances where C translations attempted to allocate extremely large uninitialized global arrays, sometimes exceeding 16~GB of memory. This pattern is tied to the source program: when the source language uses large global data structures, all three models tend to reproduce the same pattern in the target language without adapting to standard memory constraints. The rate is Low across all models because only a small subset of problems in the benchmarks involves such large global allocations. Without explicit prompt guidance to constrain memory usage, all models exhibit this behavior when the source code does.

\noindent\textbf{Uninvoked entry-point logic.}
In Avatar and CodeNet, evaluation relies on capturing standard output (stdout). We found cases where the LLM correctly implemented the required logic within a function or class but failed to generate the main execution block to invoke it. The program compiles and runs successfully but produces no output, leading to a test mismatch. All three models exhibit this at a Low rate, but occurrence is dataset-dependent: it arises specifically in problems where a main invocation block must be explicitly generated, such as in Avatar Java-to-Python translations. This issue is reliably caught and corrected by a simple rule-based post-processing check on whether the entry point is invoked. This shows that a lightweight post-processing step can recover a non-trivial number of otherwise penalized translations without modifying the model or prompt.

\subsection{Evaluation Configuration Errors in Test Suites}

\noindent\textbf{Faulty PyTest-to-JUnit conversion.} The automated conversion of EvalPlus test suites from Python (PyTest) to Java (JUnit) introduced errors in approximately 5-10\% of EvalPlus cases (164 samples) under standard prompting. Errors arise primarily when source and target languages have type mismatches that are not accounted for in the conversion prompt. In several instances, the LLM hallucinated incorrect assertions or failed to port the test logic accurately, causing correct Java translations to fail tests because the ground-truth test case itself is flawed. When the prompt is designed to provide explicit target data types and handle type conversions, the error rate drops to Low for all three models, with no substantial difference in behavior across GPT-4o, DeepSeek-Coder, and Magicoder. If LLMs are iteratively re-prompted upon test failure, the occurrence of this issue can be reduced.

\noindent\textbf{Unrealistic timeout constraints.} We identified a systemic bias where correctly translated code failed due to rigid timeout limits inherited from the source language. Benchmarks often enforce execution time limits set for the source implementation (e.g., highly optimized C++). Consequently, valid translations in languages with higher runtime overhead, such as Python (due to interpretation) or Java (due to JVM startup)---are unfairly penalized even when they implement the correct algorithm. We suggest applying language-aware timeout scaling, where limits are adjusted based on the expected overhead of the target runtime rather than inherited from the source.

\subsection{Genuine Translation Limitations}
Unlike evaluation-induced failures, these cases reflect fundamental semantic gaps between languages that cannot be resolved by fixing the evaluation pipeline alone.

The following failures reflect inherent challenges of cross-language translation rather than evaluation pipeline errors. They are reported here for completeness but are not counted as evaluation-induced false failures.

\noindent\textbf{Missing native APIs.}
Certain source-language features lack direct equivalents in the target ecosystem. For example, Python's dynamic \texttt{eval()} function has no native counterpart in Java or C++. LLMs often attempt to reimplement such features from scratch, resulting in custom implementations that are incomplete or buggy, leading to runtime crashes. 

\noindent\textbf{Divergent API semantics.} Subtle semantic differences in operators across languages cause silent failures. A prominent example is the modulo operator (\texttt{\%}) applied to negative operands: Python retains the sign of the divisor, whereas Java and C retain the sign of the dividend. Such discrepancies result in semantic mismatches where code runs without error but produces incorrect outputs. 

\noindent\textbf{Logical and corner-case errors.}
Some failures stem from missing corner-case handling that cannot be inferred without deep knowledge of the source context. For instance, Python's \texttt{round()} function employs banker's rounding (e.g., \texttt{round(2.5) $\rightarrow$ 2}), whereas C, C++, and Java typically round away from zero. We additionally identified 4 cases in the Avatar dataset where the test cases themselves contained logical errors; these data points were filtered from our evaluation.

\balance

\section{Threats to Validity}
\label{sec:threats}

\noindent\textbf{Manual inspection scope.}
We manually inspected 150 sampled failure cases out of approximately 600 observed failures across 6,164 translations. While the sample covers multiple datasets and target languages, it may not capture the full distribution of failure patterns.

\noindent\textbf{Boundary between evaluation and translation failures.}
Some failure categories do not fall cleanly into evaluation-induced or translation errors. We distinguish three types: (1) \textit{pipeline-induced} failures that affect any model under the same evaluation setup (e.g., missing library flags, unrealistic timeout constraints); (2) \textit{model-dependent} failures whose frequency varies across LLMs (e.g., natural language contamination, excessive global memory usage, faulty PyTest-to-JUnit conversion, uninvoked entry-point logic); and (3) \textit{genuine translation limitations} that reflect the inherent difficulty of cross-language semantics (e.g., missing native APIs, divergent operator semantics, logical corner-case errors). Categories (2) and (3) are not classified as evaluation-induced; they are reported separately to avoid conflating evaluation pipeline flaws with model limitations.

\noindent\textbf{Model coverage.}
All 6,164 translations are run across GPT-4o, DeepSeek-Coder, and Magicoder. However, the manual inspection of 150 failure cases was conducted on GPT-4o outputs. Pipeline-induced failures are by definition model-independent, but the sampled inspection may not fully reflect how model-dependent failure rates distribute across DeepSeek-Coder and Magicoder.

\noindent\textbf{Annotator agreement.}
Three annotators reviewed separate, non-overlapping subsets of the 150 inspected cases. Because subsets did not overlap, inter-rater agreement could not be computed. Future work should include overlapping annotation to formally measure classification consistency.

\section{Conclusion}
This work investigates the causes of false failure verdicts in LLM-based code translation across five languages and three benchmarks, covering 6,164 translations. We demonstrate that semantically correct translations are frequently rejected due to evaluation-induced errors, specifically, missing compilation flags, incorrect linker commands, and unrealistic timeout constraints. We distinguish pipeline-induced failures, which affect any model under the same evaluation setup, from model-dependent failures that vary across LLMs. Analysis across GPT-4o, DeepSeek-Coder, and Magicoder confirms that pipeline-induced issues are not artifacts of a specific model. Future work will extend this analysis to broader software engineering tasks, incorporating more languages, datasets, and models.

\section*{Acknowledgement}
\label{acknowledgement}
\noindent This research was supported by the Fonds de recherche du Québec (Grant No.2024-NOVA346499)\citep{nova}, Natural Sciences and Engineering Research Council of Canada (NSERC) through the Alliance, Grant (Grant No.586838-23), the NSERC Discovery Grant (Grant No. RGPIN-2019-07007 and Grant No. DGECR-2019-00464), and NSERC CREATE Grant (Grant No.555406-2021). We gratefully acknowledge the support of all funding agencies.

\clearpage

\end{document}